\documentclass[twocolumn,showpacs,preprintnumbers,amsmath,amssymb]{revtex4}
\usepackage{graphicx}% Include figure files
\usepackage{dcolumn}% Align table columns on decimal point
\usepackage{bm}% bold math

\begin{document}

\preprint{}

\title{Enhanced and switchable spin Hall effect of light near the Brewster angle on reflection}% Force line breaks with \\
\author{Hailu Luo}
%\altaffiliation[ ]{}%Lines break automatically or can be forced with \\
\author{Xinxing Zhou}
\author{Weixing Shu}
\author{Shuangchun Wen}\email{scwen@hnu.cn}
\author{Dianyuan Fan}
%%\email{Second.Author@institution.edu}
\affiliation{Key Laboratory for Micro/Nano Opto-Electronic Devices
of Ministry of Education, College of Information Science and
Engineering, Hunan University, Changsha 410082, People's Republic of
China}
\date{\today}% It is always \today, today,

\begin{abstract}
We reveal an enhanced and switchable spin Hall effect (SHE) of light
near Brewster angle on reflection both theoretically and
experimentally. The obtained spin-dependent splitting reaches 3200nm
near Brewster angle, 50 times larger than the previous reported
values in refraction. We find that the amplifying factor in week
measurement is not a constant which is significantly different from
that in refraction. As an analogy of SHE in electronic system, a
switchable spin accumulation in SHE of light is detected. We were
able to switch the direction of the spin accumulations by slightly
adjusting the incident angle.

\end{abstract}


%Use showkeys class option if keyword
\maketitle

\section{Introduction}\label{SecI}
The spin Hall effect (SHE) of light can be regarded as a direct
optical analogy of SHE in electronic system where the spin electrons
and electric potential are replaced by spin photons and refractive
index gradient, respectively~\cite{Onoda2004,Bliokh2006,Hosten2008}.
Recently, the SHE of light has been intensively investigated in
different physical systems, such as high-energy
physics~\cite{Gosselin2007}, plasmonics~\cite{Gorodetski2008},
optical
physics~\cite{Bliokh2008,Haefner2009,Aiello2009,Herrera2010}, and
semiconductor physics~\cite{Menard2010}. The SHE of light is
generally believed as a result of an effective spin-orbital
interaction, which describes the mutual influence of the spin
(polarization) and trajectory of the light beam. In general, the
spin-dependent splitting in these physics systems is limited by a
fraction of the wavelength, and therefore it is disadvantage for
potential application to nano-photonic devices.

The SHE in electronic system offer an effective way to manipulate
the spin particles, and open a promising way to potential
applications in semiconductor spintronic
devices~\cite{Murakami2003,Sinova2004,Wunderlich2005}. The
generation and manipulation of spin-polarized electrons in
semiconductors define the main challenges of spin-based
electronics~\cite{Wolf2001}. In semiconductor systems, the spin
accumulation can be switched by altering the directions of external
magnetic field~\cite{Kimura2007,Mihaly2008}. By rotating the
polarization plane of the exciting light, the directions of spin
current can be switched in a semiconductor
micro-cavity~\cite{Kavokin2005,Leyder2007}. Now a question arises:
Whether there exists a similar phenomenon in SHE of light? In this
paper, we want to reveal an enhanced and switchable SHE of light
near Brewster angle on reflection. The SHE of light has been studied
in reflection both in theory~\cite{Bliokh2007,Aiello2008,Luo2009}
and in experiment~\cite{Qin2009}. However, the developed paraxial
propagation model cannot be applied and the experimental evidence is
still absent for describing the SHE of light near Brewster angle.

The paper is organized as follows. First, we develop a general
propagation model to describe the SHE of light near Brewster angle
on reflection. Next, we attempt to reveal the enhanced SHE of light
in theory and detect the large spin-dependent splitting in
experiment via week measurements. The large spin-dependent splitting
is found to be attributed to the large ratio between the Fresnel
reflection coefficients near Brewster angle. Finally, we want to
explore the switchable SHE of light. We demonstrate that the
transverse displacements can be tuned to either a negative, or a
positive value, or even zero, by slightly adjusting the incident
angle. The underlying secret can be interpreted from that the
horizontal field component changes its phase across the Brewster
angle. As an analogy of SHE in electronic system, the spin
accumulations can be switched in the SHE of light.

\section{General propagation model}\label{SecII}
\begin{figure}
\centerline{\includegraphics[width=8cm]{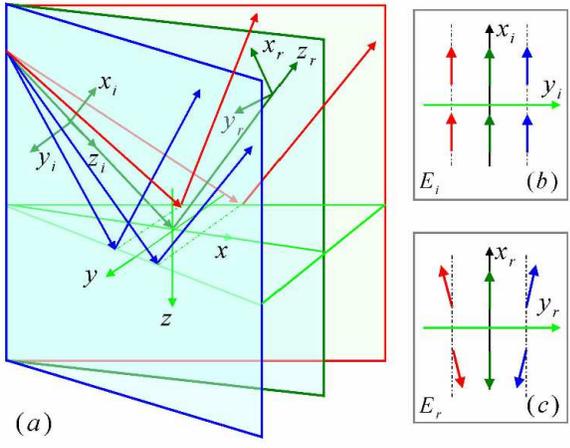}}
\caption{\label{Fig1}(color online) (a) Plane-wave components in
four quadrant acquire different polarization rotations upon
reflection to satisfy transversality. The polarizations associated
with the angular spectrum components in incidence (b) experience
different rotations in reflection (c).}
\end{figure}

We first develop a general propagation model to describe the SHE of
light near the Brewster angle on reflection. The $z$ axis of the
laboratory Cartesian frame ($x,y,z$) is normal to the air-prism
interface. We use the coordinate frames ($x_i,y_i,z_i$) and
($x_r,y_r,z_r$) to denote incident and reflection, respectively
[Fig.~\ref{Fig1}(a)]. In the spin basis set, the incident angular
spectrum can be written as:
\begin{equation}
\tilde{\mathbf{E}}_i^H=\frac{1}{\sqrt{2}}(\tilde{\mathbf{E}}_{i+}+\tilde{\mathbf{E}}_{i-})\label{SBH},
\end{equation}
\begin{equation}
\tilde{\mathbf{E}}_i^V=\frac{1}{\sqrt{2}}i(\tilde{\mathbf{E}}_{i-}-\tilde{\mathbf{E}}_{i+})\label{SBV}.
\end{equation}
Here, $H$ and $V$ represent horizontal and vertical polarizations,
respectively.
$\tilde{\mathbf{E}}_{i+}=(\mathbf{e}_{ix}+i\mathbf{e}_{iy})\tilde{E}_{i}/\sqrt{2}$
and
$\tilde{\mathbf{E}}_{i-}=(\mathbf{e}_{ix}-i\mathbf{e}_{iy})\tilde{E}_{i}/\sqrt{2}$
denote the left and right circularly polarized (spin) components,
respectively. We consider the incident beam whose angular spectrum
with a Gaussian distribution:
\begin{equation}
\tilde{E}_{i}=\frac{w_0}{\sqrt{2\pi}}\exp\left[-\frac{w_0^2(k_{ix}^2+k_{iy}^2)}{4}\right]\label{asi},
\end{equation}
where $w_0$ is the beam waist. The complex amplitude for the
reflected beam can be conveniently expressed as
\begin{eqnarray}
\mathbf{E}_r(x_r,y_r,z_r )&=&\int d k_{rx}dk_{ry}
\tilde{\mathbf{E}_r}(k_{rx},k_{ry})\nonumber\\
&&\times\exp [i(k_{rx}x_r+k_{ry}y_r+ k_{rz} z_r)],\label{apr}
\end{eqnarray}
where $k_{rz}=\sqrt{k_r^2-(k_{rx}^2+k_{ry}^2)}$ and
$\tilde{\mathbf{E}_r}(k_{rx},k_{ry})$ is the reflected angular
spectrum.

The reflected angular spectrum is related to the boundary
distribution of the electric field by means of the
relation~\cite{Bliokh2006}
\begin{eqnarray}
\left[\begin{array}{cc}
\tilde{E}_r^H\\
\tilde{E}_r^V
\end{array}\right]
=\left[
\begin{array}{cc}
r_p&\frac{k_{ry} (r_p+r_s) \cot\theta_i}{k_0} \\
-\frac{k_{ry} (r_p+r_s)\cot\theta_i}{k_0} & r_s
\end{array}
\right]\left[\begin{array}{cc}
\tilde{E}_i^H\\
\tilde{E}_i^V
\end{array}\right]\label{matrixr},
\end{eqnarray}
where $r_p$ and $r_s$ denote the Fresnel reflection coefficients for
parallel and perpendicular polarizations, respectively. By making
use of Taylor series expansion based on the arbitrary angular
spectrum component, $r_p$ and $r_s$ can be expanded as a polynomial
of $k_{ix}$:
\begin{eqnarray}
r_{p,s}(k_{ix})&=&r_{p,s}(k_{ix}=0)+k_{ix}\left[\frac{\partial
r_{p,s}(k_{ix})}{\partial
k_{ix}}\right]_{k_{ix}=0}\nonumber\\
&&+\sum_{j=2}^{N}\frac{k_{ix}^N}{j!}\left[\frac{\partial^j
r_{p,s}(k_{ix})}{\partial k_{ix}^j}\right]_{k_{ix}=0}\label{LMD}.
\end{eqnarray}
The reflection coefficient changes its sign across the Brewster
angle, which means the electric field reverses its directions
[Fig.~\ref{Fig1}(b) and~\ref{Fig1}(c)]. The polarizations associated
with the angular spectrum components experience different rotations
in order to satisfy the boundary condition after reflection.

In the spin basis set, the reflected angular spectrum can be written
as:
\begin{equation}
\tilde{\mathbf{E}}_r^H=\frac{1}{\sqrt{2}}(\tilde{\mathbf{E}}_{r+}+\tilde{\mathbf{E}}_{r-})\label{JRH},
\end{equation}
\begin{equation}
\tilde{\mathbf{E}}_r^V=\frac{1}{\sqrt{2}}i(\tilde{\mathbf{E}}_{r-}-\tilde{\mathbf{E}}_{r+})\label{JRV}.
\end{equation}
We consider the incident Gaussian beam with $H$ polarization. From
the boundary condition, we obtain $k_{rx}=-k_{ix}$ and $k_{ry}=
k_{iy}$. In fact, after the incident angular spectrum is known,
Eq.~(\ref{apr}) together with Eqs.~(\ref{asi})-(\ref{JRV}) provides
the general expression of the reflected field:
\begin{eqnarray}
\mathbf{E}_{r\pm}^H&=&\frac{r_p(\mathbf{e}_{rx}\pm
i\mathbf{e}_{ry})}{\sqrt{\pi}w_0}\frac{z_R}{z_R+iz_r}
\exp\left[-\frac{k_0}{2}\frac{x_{r}^2+y_{r}^2}{z_R+iz_r}\right]\nonumber\\
&&\times\bigg[r_p-\frac{ix}{z_R+iz_r}\frac{\partial
r_{p}}{\partial\theta_i}\pm\frac{y}{z_R+iz_r}(r_p+r_s)\nonumber\\
&&\pm\frac{ixy}{(z_R+iz_r)^2}\left(\frac{\partial
r_{p}}{\partial\theta_i}+\frac{\partial
r_{s}}{\partial\theta_i}\right)\bigg]\exp(ik_rz_r)\label{HPR},
\end{eqnarray}
where $z_R=k_0w_0^2/2$ is the Rayleigh lengths. Our analysis is
confined to the first order in Taylor series expansion of Fresnel
reflection coefficients.

\section{Spin Hall effect of Light}\label{SecIII}
We now determine the sin-dependent splitting of field centroid. At
any given plane $z_a=\text{const.}$, the transverse displacement of
field centroid compared to the geometrical-optics prediction is
given by
\begin{equation}
\delta_{\pm}^H= \frac{\int \int y_r I_{\pm}^H(x_r,y_r,z_r)
\text{d}x_r \text{d}y_r}{\int \int I_{\pm}^H(x_r,y_r,z_r)
\text{d}x_r \text{d}y_r}.\label{centroid}
\end{equation}
The intensity distribution of beam is closely linked to the
longitudinal momentum currents
$I(x_r,y_r,z_r)\propto\mathbf{p}_r\cdot \mathbf{e}_{rz}$. The
time-averaged linear momentum density associated with the
electromagnetic field can be shown to be
$\mathbf{p}_{r}\propto\mathrm{Re}[\mathbf{E}_{r}
\times\mathbf{H}_{r}^\ast]$, where the magnetic field can be
obtained by $\mathbf{H}_{r}=-ik_{r}^{-1}
\nabla\times\mathbf{E}_{r}$.

\begin{figure}
\centerline{\includegraphics[width=8cm]{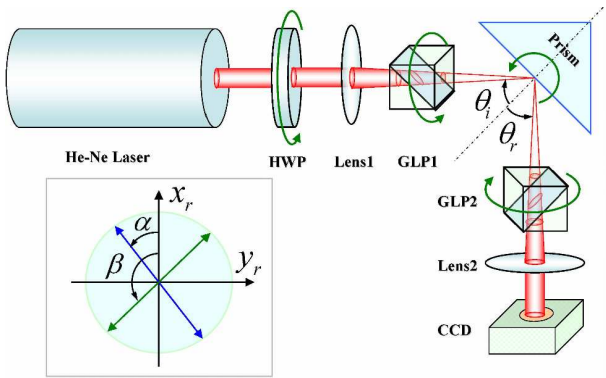}}
\caption{\label{Fig2} (Color online) (a) Experimental setup for
characterizing the SHE of light in reflection near Brewster angle.
Prism with refractive index $n=1.515$ (BK7 at $632.8$ nm); Lens1 and
Lens2, lens with effective focal lengths $50$ mm and $250$ mm,
respectively; HWP, half-wave plate (for adjusting the intensity);
GLP1 and GLP2, Glan Laser polarizers; CCD, charge-coupled device
(Coherent LaserCam HR); The light source is a $17$-mW linearly
polarized He-Ne laser at $632.8$ nm (Thorlabs HRP170). The inset
clarify Glan Laser polarizers whose axis make angles $\alpha$ and
$\beta$ with $x_r$.}
\end{figure}

To detect the displacements, we use the signal enhancement
technique~\cite{Hosten2008} known from weak
measurements~\cite{Aharonov1988,Ritchie1991}.  In principle, this
enhancement mechanism of this setup can be perfectly presented in a
classical description~\cite{Aiello2008}. Figure~\ref{Fig2}
illustrates the experimental setup. A Gaussian beam generated by a
He-Ne laser passes through a short focal length lens (Lens1) and a
polarizer (GLP1) to produce an initially polarized focused beam.
When the beam impinges onto the prism interface, the SHE of light
generates. The prism was mounted to a rotation stage which allows
for precise control of the incident angle $\theta_i$. The incident
beam is preselected in the $H$ polarization state ($\alpha=0$) by
GLP1, and then postselected ($\beta=\pi/2+\Delta$) by GLP2 in the
polarization state with
\begin{equation}
\mathbf{V}=\sin\Delta\mathbf{e}_{rx}+\cos\Delta\mathbf{e}_{ry}.
\end{equation}
In our measurement, we chose $\Delta=2\pm0.04^{\circ}$. Note that
the interesting cross polarization effect can be observed as
$\Delta=0$~\cite{Aiello2009b}. As the reflected beam of light splits
by several wavelengths, the intensity distribution on the prism
interface is nearly unchanged. After the second polarizer GLP2, the
two splitting components interfere, and produce a field
redistribution whose centroid is significantly amplified. We use a
CCD to measure the amplified displacement after a long focal length
lens (Lens2).

\begin{figure}
\centerline{\includegraphics[width=8cm]{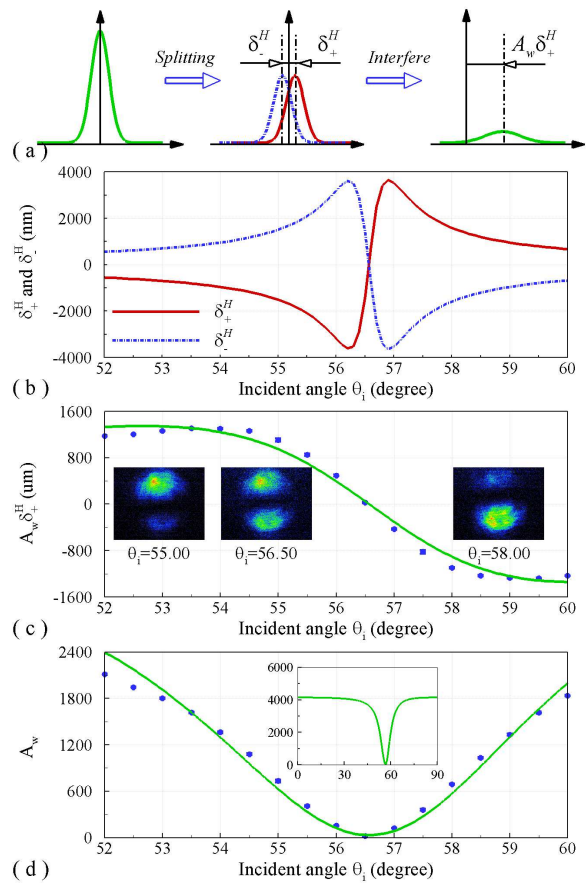}}
\caption{\label{Fig3} (Color online) (a) Presection and
postselection of polarization give rise to an interference in the
CCD, shifting it to its final centroid position proportional to
$A_w=\delta_w/\delta_+^H$. (b) Theoretical spin-dependent transverse
splitting of spin components at the prism interface. (c) Theoretical
and experimental results for amplifying displacements $\delta_w$.
Insets show the measured field distribution. (d) Theoretical and
experimental results for amplifying factor $A_w$ in the week
measurement. Inset presents a full view.}
\end{figure}

The week measurement of SHE of light is schematically shown in
Fig.~\ref{Fig3}(a). The theoretical transverse displacements given
in Eq.~(\ref{centroid}) show that  the two opposite spin components
would have opposite tendency versus $\theta_i$ [Fig.~\ref{Fig3}(b)].
It indicates that the SHE of light can be greatly enhanced near the
Brewster angle. We obtain the value of spin-dependent splitting
$3200$nm at $\theta_i=56^{\circ}$ and 50 times larger than the
previous reported values of refraction~\cite{Hosten2008}. The
relevant amplitude of the reflected field at the plane of $z_r$ can
be obtained as $\mathbf{V}\cdot\mathbf{E}_{r}^H$. The amplified
displacement of field centroid $\delta_{w}$ at the CCD is much
larger than the original displacement $|\delta_{\pm}^H|$.
Calculation of the centroid of the distribution of
$\mathbf{V}\cdot\mathbf{E}_{r}^H$ yields the amplifying factor
$A_w=\delta_{w}/\delta_{+}^H$. Our experimental results for the
amplified displacement $\delta_{w}$ versus the incident angle
$\theta_i$ are reported in Fig.~\ref{Fig3}(c). We measure the
displacements every $0.5^{\circ}$ from $52^{\circ}$ to $60^{\circ}$.
The measure values allow for calculating the original displacement
caused by SHE of light. The solid lines represent the theoretical
predictions. It should be noted that the amplifying factor in week
measurements is always the same in refraction~\cite{Hosten2008}.
However, it presents a valley near Brewster angle on reflection
[Fig.~\ref{Fig3}(d)]. The experimental results are in good agreement
with the theory without using parameter fit.

From Eq.~(\ref{HPR}), we know that the transverse displacements are
related to the ratio between the Fresnel transmission coefficients
$r_p$ and $r_s$. The reflection coefficient of horizontal
polarization $r_p$ vanishes at exactly the Brewster angle, and
changes its sign across the angle. Hence, the large spin-dependent
splitting in SHE of light is attributed to the large ratio of
$r_s/r_p$ near the Brewster angle. On the contrary, a small ratio of
$r_s/r_p$ would greatly suppress the SHE of light. It should be
mentioned that a large value of $\partial r_{p}/\partial\theta_i$
near Brewster angle will lead to large Goos-Hanchen
shifts~\cite{Lai2002} and angular shifts~\cite{Merano2009}. It
should be noted that the horizontal component of electric field
alters its phase, however the vertical component does not. As a
result, the phase difference $\arg[r_s]-\arg[r_p]$ experiences a
variation $\pi$, and the spin accumulation would reverse its
directions accordingly. Due to the reversed spin-dependent
splitting, the spin accumulation can be switched by slightly
adjusting the incident angle.

The SHE of light may open new opportunities for manipulating photon
spin and developing new generation of all-optical devices as
counterpart of recently presented spintronics
devices~\cite{Hosten2008,Wolf2001}. It should be mentioned that the
spatial separation of the spin components is very small in the
refraction. Hence, it is disadvantage for potential application to
nano-photonic devices. In refraction~\cite{Hosten2008} and photon
tunneling~\cite{Luo2010} the reversed spin accumulation requires the
reversed refractive index gradient. As shown in above, the
transverse displacements can be tuned to either a negative, or a
positive value, or even zero, by just adjusting the incident angle.
Hence, our scheme provide more flexibility for switching the
direction of the spin accumulations. These interesting phenomena
open a promising way to some potential applications in spin-based
nano-photonic devices. Because of the close similarity of Brewster
angle in optical physics, condensed matter~\cite{Khodas2004}, and
plasmonics~\cite{Gordon2006}, by properly facilitating the
reflection near Brewster angle, the SHE may be effectively modulated
in these physical systems.

\section{Conclusions}
In conclusion, we have revealed an enhanced and switchable
spin-dependent splitting near Brewster angle on reflection. The
detected spin-dependent splitting reaches 3200nm near Brewster
angle, and 50 times larger than the previous reported values in
refraction. We have found that the amplifying factor is not a
constant which is significantly different from that in the
refraction case. The enhanced spin-dependent splitting is found to
be attributed to the large ratio between the Fresnel reflection
coefficients near Brewster angle. As an analogy of SHE in electronic
system, the switchable SHE of light has been detected, which can be
interpreted from the inversion of horizontal electric field vector
across the Brewster angle. We were able to switch the directions of
the spin accumulation, by slightly adjusting the incident angle near
Brewster angle. These findings provide a novel pathway for
modulating the SHE of light, and thereby open the possibility for
developing new nano-photonic devices.

\begin{acknowledgements}
This research was partially supported by the National Natural
Science Foundation of China (Grants Nos. 61025024, 11074068, and
10904036).
\end{acknowledgements}

\end{document}